\documentclass[preprint,12pt,authoryear]{elsarticle}

\usepackage{amssymb}
\usepackage{amsmath}
\usepackage{xcolor}
\usepackage{lscape}
\usepackage[ruled,vlined]{algorithm2e}   

\begin{document}

\begin{frontmatter}



\title{Reinforcement learning meets bioprocess control through behaviour cloning: Real-world deployment in an industrial photobioreactor} 


\author[First]{Juan D. Gil}\corref{corr}\ead{juandiego.gil@ual.es} 
\cortext[corr]{Corresponding author}
\author[Second]{Ehecatl Antonio Del Rio Chanona}\ead{a.del-rio-chanona@imperial.ac.uk } 
\author[First]{José L. Guzmán}\ead{joguzman@ual.es} 
\author[First]{Manuel Berenguel}\ead{beren@ual.es}

\affiliation[First]{organization={Centro Mixto CIESOL, ceia3, Department of Informatics, Universidad de Almería},
            addressline={Ctra. Sacramento s/n}, 
            city={Almería},
            postcode={04120}, 
            country={Spain}}

\affiliation[Second]{organization={Sargent Centre for Process Systems Engineering},
            addressline={Imperial College London}, 
            postcode={SW7 2AZ}, 
            state={London},
            country={UK}}

\begin{abstract}
The inherent complexity of living cells as production units creates major challenges for maintaining stable and optimal bioprocess conditions, especially in open Photobioreactors (PBRs) exposed to fluctuating environments. To address this, we propose a Reinforcement Learning (RL) control approach, combined with Behavior Cloning (BC), for pH regulation in open PBR systems. This represents, to the best of our knowledge, the first application of an RL-based control strategy to such a nonlinear and disturbance-prone bioprocess. Our method begins with an offline training stage in which the RL agent learns from trajectories generated by a nominal Proportional-Integral-Derivative (PID) controller, without direct interaction with the real system. This is followed by a daily online fine-tuning phase, enabling adaptation to evolving process dynamics and stronger rejection of fast, transient disturbances. This hybrid offline–online strategy allows deployment of an adaptive control policy capable of handling the inherent nonlinearities and external perturbations in open PBRs. Simulation studies highlight the advantages of our method: the Integral of Absolute Error (IAE) was reduced by 8~\% compared to PID control and by 5~\% relative to standard off-policy RL. Moreover, control effort decreased substantially—by 54~\% compared to PID and 7~\% compared to standard RL—an important factor for minimizing operational costs. Finally, an 8-day experimental validation under varying environmental conditions confirmed the robustness and reliability of the proposed approach. Overall, this work demonstrates the potential of RL-based methods for bioprocess control and paves the way for their broader application to other nonlinear, disturbance-prone systems.
\end{abstract}








\end{frontmatter}



\section{Introduction}

The importance of effective bioprocess control cannot be overstated. Unlike conventional chemical processes, where the reactor is the primary unit of control, bioprocesses rely on living cells as the actual manufacturing units. These cells are complex, autonomous systems with internal regulatory mechanisms and are heterogeneously distributed within the bioreactor. This introduces significant challenges, as the micro-scale dynamics of individual cells cannot be directly manipulated through macro-scale control variables \citep{luo2021bioprocess}. Maintaining stable environmental conditions—such as nutrient levels, pH, temperature, and dissolved oxygen—is essential to support cellular proliferation and productivity, yet these variables are in constant flux requiring the implementation of advanced control systems \citep{liu2024model}.

Microalgae-based systems exemplify these challenges. Microalgae are photosynthetic microorganisms capable of thriving under adverse conditions by converting solar energy and carbon-based compounds such as CO$_2$ into biomass, while releasing oxygen \citep{tarafdar2023environmental}. Their growth depends on the availability of essential nutrients like carbon, nitrogen, and phosphorus. CO$_2$ is typically supplied via injection, serving both as a carbon source and a pH buffer—making pH control one of the most critical system variables. Nitrogen and phosphorus can be supplied directly or derived from the medium, particularly when wastewater is used. These biological and operational complexities underscore the need for advanced, automated control strategies to ensure process stability and regulatory compliance \citep{luo2021bioprocess, wang2022does, GuzACC2025}.

Focusing on pH, this variable stands out as one of the most critical, as it directly influences the solubility and availability of both CO$_2$ and nutrients, significantly affecting the metabolism of microalgae \citep{juneja2013effects,nordio2023influence}. Photosynthesis itself induces constant pH fluctuations, further complicating its control. Typically, on/off control systems are employed \citep{rodriguez2021modelling}, which fail to capture the system’s dynamic behavior and external disturbances. Additionally, simple Proportional, Integral, and Derivative (PID) controllers with fixed parameters have been proposed as well \citep{fernandez2010modelling,isiramen2022improving}, yet their performance is often insufficient due to the nonlinearities, disturbances, and time-varying dynamics inherent to the process. The reliance on such control strategies largely stems from the challenges associated with developing accurate models that fully represent the complex process dynamics \citep{guzman2021modelling,GuzACC2025}.

Given the modeling challenges inherent in these types of processes, robust and adaptive control techniques have arisen in the literature. 
Several studies have addressed this issue from different perspectives. For instance, a robust adaptative feedforward tracking control was presented by \cite{schaum2017robust}. 
\cite{feudjio2021experimental} explored the use of Extremum Seeking Control~(ESC) to drive the productivity of a continuous photobiorreactor (PBR) to optimal or suboptimal setpoints. 
\cite{amaro2023adaptive} introduced an adaptative fuzzy strategy for the nominal model of a generalized Model Predictive Control (MPC) in charge of regulating the pH. In the work by \cite{caparroz2024novel}, an adaptive model based on a regression tree was presented, capable of predicting pH under varying operational conditions. This tree was subsequently used for the development and implementation of an adaptive control strategy based on a PID controller \citep{hagglund2024give}. In another study, \cite{caparroz2023control} explored the use of a Model Reference Adaptive Control (MRAC) strategy for pH regulation. This strategy was later hybridized with a PID controller \citep{caparroz2025hybrid}. Moreover, in \cite{caparroz2025Relay}, a new relay-based autotuning technique based on PID controllers that applies the relay signal to the setpoint was recently designed and experimentally validated for pH control in semi-industrial open PBR. However, such adaptive strategies still face challenges related to the need for nominal models and their limited generalization capability in the face of highly nonlinear and time-varying dynamics, as typically observed in PBR operating over extended periods.

In light of the limitations of adaptive approaches, data-driven learning techniques have emerged as a more flexible alternative. One of the first efforts in this direction was presented by \citet{pataro2023learning}, who developed an MPC strategy combined with an oracle function that learns online from data to adjust the uncertainties of the nominal model used internally by the MPC. This strategy yielded good results in terms of adaptability and control across dynamic variations induced by different culture media. Nonetheless, despite its performance, this approach still relies on explicit control structures and a certain degree of prior knowledge about the system dynamics. In this context, Reinforcement Learning (RL) techniques represent a further step toward greater autonomy and adaptability in control systems. Unlike MPC, which optimizes decisions based on a predictive model, RL directly learns a control policy through interaction with the environment or from historical data, allowing it to continuously adjust its behavior without requiring an explicit model representation \citep{petsagkourakis2020reinforcement}.

Despite its potential, RL still faces important challenges for control purposes \citep{bucsoniu2018reinforcement}, which are specially relevant in the bioprocess engineering field. A key limitation is the data-intensive nature of training, which typically requires either large volumes of process data or accurate simulation models—reintroducing the modeling bottleneck. Moreover, RL agents often rely on machine learning models such as Artificial Neural Networks (ANNs) to parameterize both the value function and control policy, making them highly dependent on data quality and quantity. In this context, the distinction between \textit{on-policy} and \textit{off-policy} learning becomes critical. While \textit{on-policy} methods require the agent to collect new data by acting according to its current policy \citep{sachio2021simultaneous}, \textit{off-policy} approaches allow the agent to learn from data previously collected under different  policies \citep{Haiting2025}. This capability is especially valuable in PBR applications, where online experimentation is costly and risky. 

\textit{Off-policy} RL seeks to learn policies that can be deployed in real-world settings without further exploration—an essential advantage when direct interaction with the environment is impractical or risky \citep{levine2020offline}. Leveraging this property, several authors have designed model-free controllers for a variety of domains. For example, \citet{deng2023offline} trained an offline RL agent with historical data from a steel plant and evaluated its performance in a simulator of the system. Similarly, \citet{schepers2022autonomous} applied offline RL to regulate indoor temperatures in buildings, while \citet{blad2022data} explored a multi-agent offline RL approach for the same task. Although these works demonstrate the potential of offline RL, they rely exclusively on fixed datasets and are validated only in simulation. Consequently, the resulting policies cannot adapt when the system dynamics shift—a limitation that becomes acute in bioprocesses and other time-varying environments.

Beyond reusing open-loop operational data, a key direction in \textit{off-policy} RL is to exploit experiences from expert systems—nominal controllers designed with established control theory or even human operator actions.  
A widely used technique for this purpose is \textit{Behavior Cloning} (BC), a supervised-learning method that trains the policy to imitate optimal behaviors contained in an offline dataset \citep{levine2020offline}.  
Because the data reflects stable and highly performing behaviors, BC directly transfers this domain knowledge to the agent, mitigating a long-standing limitation of RL algorithms \citep{bucsoniu2018reinforcement}, while avoiding the risks and costs of online exploration. Even though this approach shows great potential, only a few works have explored it in the literature. One example was presented by \citet{seo2025implementation}, who trained an RL agent for pressure control in a crude distillation unit using data derived from human operator decisions, then improved it through online retraining. More recently, \citet{Haiting2025} combined an offline RL policy—trained on closed-loop trajectories generated by an MPC controller—with an online fine-tuning phase that adapts to time-varying dynamics of the system. The objective was to optimize the production of an in-silico semi-batch bioprocess. Such hybrid approaches are particularly attractive in domains like bioprocess engineering, where system dynamics can drift, and purely offline policies quickly become sub-optimal. However, most of these strategies have limited applicability to real-world process control so far, as the majority of studies have been restricted to simulations or relatively simple systems. This highlights a gap in the existing literature that must be addressed in order to advance the application of Machine Learning~(ML) techniques to the control of complex systems.

Building on the ideas of these hybrid methodologies, and motivated by the inherently dynamic and non-linear behavior of bioprocesses such as those involving microalgae, this work proposes an \textit{off-policy} reinforcement learning control system for pH regulation in microalgae PBR. Specifically, the proposed model-free control algorithm is based on a Deep Deterministic Policy Gradient (DDPG) agent \citep{nieto2025control, rajasekhar2025exploring}. This strategy learns from historical experiences generated by conventional controllers, such as PID regulators, without requiring direct interaction with the real system. Once deployed, the agent can continue its training periodically with new data, which enables it to adapt to the time-varying dynamics and disturbances effects. The most significant achievement of this work is demonstrating, for the first time to our knowledge, the successful real-world deployment of a RL–based approach for controlling a complex, highly nonlinear, and multi-disturbed system. Moreover, compared to previous studies, the proposed approach stands out as:

\begin{enumerate}
    \item A Partially Observable Markov Decision Process (POMDP) formulation is developed for a demonstration-scale bioprocess, explicitly accounting for fast disturbances such as solar irradiance fluctuations, air injections, and dilution rate variations. This observation space allows the agent to incorporate an intrinsic feedforward control action, along with various variables that help it navigate the partial observability of the process.

    \item From a reinforcement learning perspective, a comprehensive methodology is proposed to develop a hybrid offline–online RL-based control strategy. First, the agent is trained offline using experience collected during the operation of a basic nominal PID-type controller. It is worth noting that PID controllers represent a form of nominal control commonly present in virtually all industrial processes \citep{hagglund2024give}. Second, an online fine-tuning stage is introduced to adapt to evolving process dynamics and enhance the rejection of fast, transient disturbances.

    \item The proposed approach is experimentally validated in an open PBR at demonstration-scale under industrially relevant conditions, during eight days of operation at the Solar Energy Research Center (CIESOL) facilities located at the Andalusian Institute for Agricultural, Fisheries, Food and Organic Production Research and Training (IFAPA) centre close to the University of Almería (UAL). To the best of our knowledge, this represents the first experimental validation of an RL-based control strategy in a bioprocess with highly dynamic and multi-disturbance characteristics—providing a relevant contribution to the field of advanced bioprocess control.
\end{enumerate}

The paper is organized as follows: Section~\ref{section_2} presents the main materials used in this study, along with the theoretical background on reinforcement learning. Section~\ref{section3} details the proposed methodology. Section~\ref{section4} discusses the main results obtained from applying the proposed approach to the PBR systems. Finally, Section~\ref{section5} provides the main conclusions.
    
\section{Material and methods}
\label{section_2}

\subsection{System Overview and Control Problem}
\subsubsection{System Description}
\begin{figure}[ht]
\centering
  \includegraphics[width=0.65\linewidth]{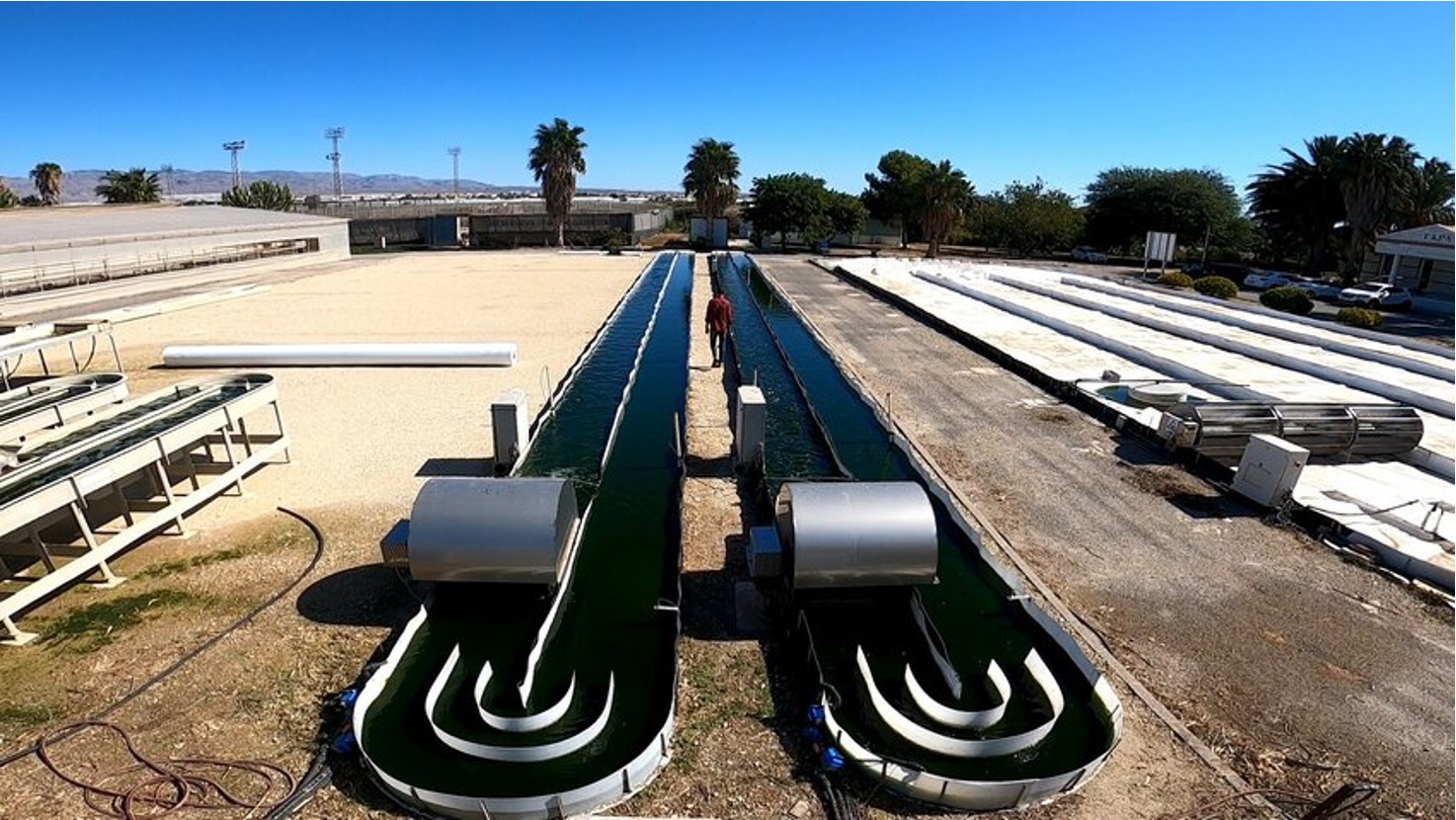}\\
  \caption{Real open PBR facilities, in Almería, Spain. The two reactors have identical characteristics. The one used in this work is the reactor on the left.}\label{fig1}
\end{figure}

\begin{figure}[ht]
\centering
  \includegraphics[width=0.68\linewidth]{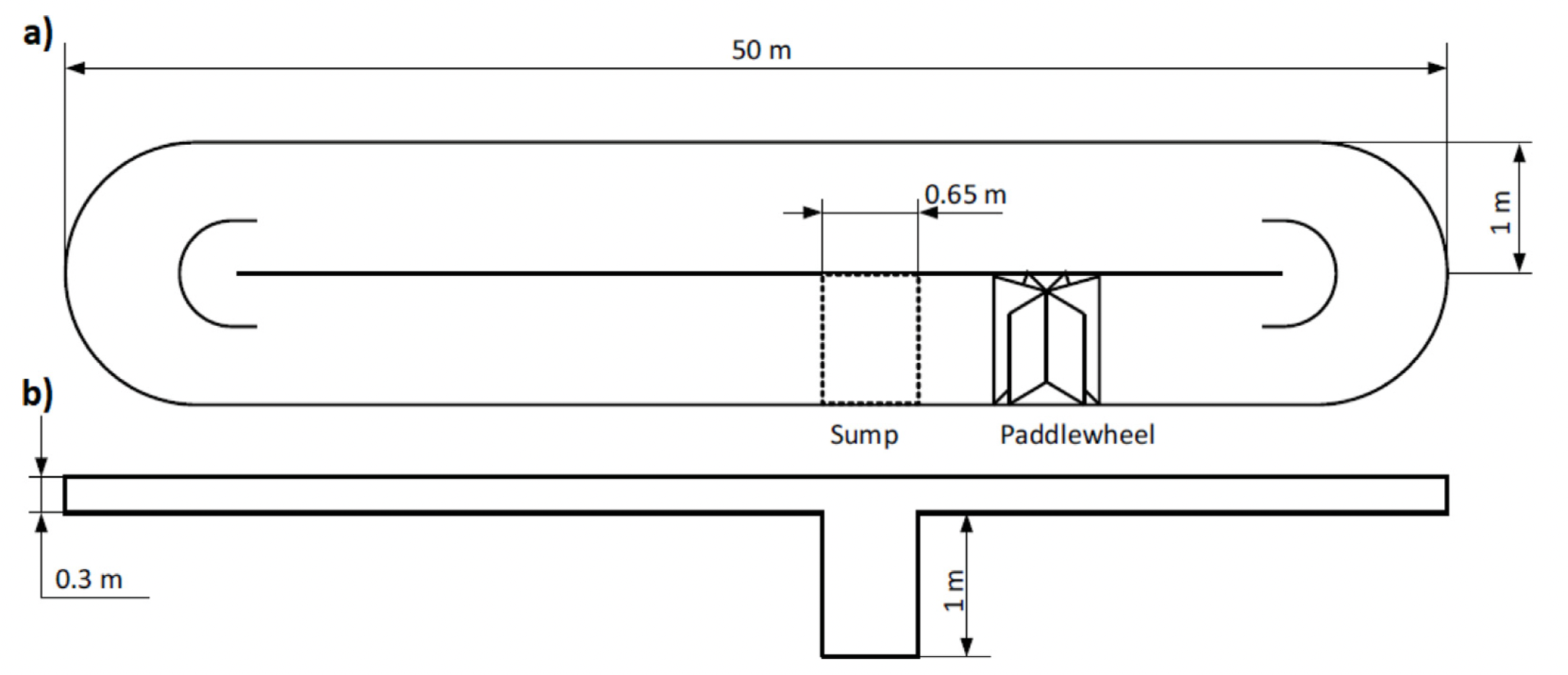}\\
  \caption{PBR physical scheme, (a) top view, (b) side view.}\label{fig2}
\end{figure}

The \textit{raceway} PBR used in this study has a surface area of 80 m$^2$ and is located at the IFAPA center of the Junta de Andalucía, near the UAL (see Fig.~\ref{fig1}). This open PBR is primarily used for microalgae biomass production. It consists of two channels, each measuring 50 meters in length, 1 meter in width, and 0.3 meters in depth (see Fig.~\ref{fig2}). Mixing and circulation are achieved via a paddlewheel system with a diameter of 1.2 meters and equipped with eight blades. CO$_2$ is injected 1 meter deep into a sump measuring 0.65 by 1 meter, situated 1.8 meters upstream from the paddlewheel. The paddlewheel’s role is to ensure continuous recirculation, flow, and homogenization of the culture medium. Flow velocity is regulated by a frequency inverter controlling the paddlewheel’s rotation speed, maintaining a constant velocity during operation.

The system is equipped with comprehensive instrumentation that enables real-time data collection, recording various process variables every second, such as pH, dissolved oxygen, culture temperature and level, as well as environmental parameters including solar radiation, air temperature, wind speed, and relative humidity, among others. Measurements of pH and Dissolved Oxygen (DO) are taken at two critical points: just after the sump and at the end of the channel, right before the paddlewheel. The latter point poses the greatest control challenge, as it is farther from the CO$_2$ injection area and, therefore, is the main focus of the regulation strategies implemented in the system.

\subsubsection{Control problem statement}

To analyze the control problem in PBR systems, it is essential to understand the physical and biological principles governing pH dynamics within the reactors. In microalgae cultivation using freshwater media supplemented with fertilizers (as in the PBR employed in this study—see \cite{morillas2020year} for details), the pH of the culture medium is primarily influenced by CO$_2$ uptake during photosynthesis. During this process, microalgae consume CO$_2$, producing O$_2$ and causing an increase in pH. In contrast, dissolved CO$_2$, which is injected as a control signal, reacts to form carbonic acid, which lowers the pH. Additionally, solar irradiance (I) serves as the principal energy source determining the rate of photosynthesis, directly affecting microalgae growth and thus influencing pH dynamics. Other important variables impacting photosynthesis include DO and culture medium temperature (T), both contributing to the overall biological activity and consequently affecting pH behavior in the system \citep{nordio2024abaco}.

\begin{figure}[ht]
\centering
  \includegraphics[width=0.75\linewidth, clip, trim=10cm 10cm 10cm 2cm]{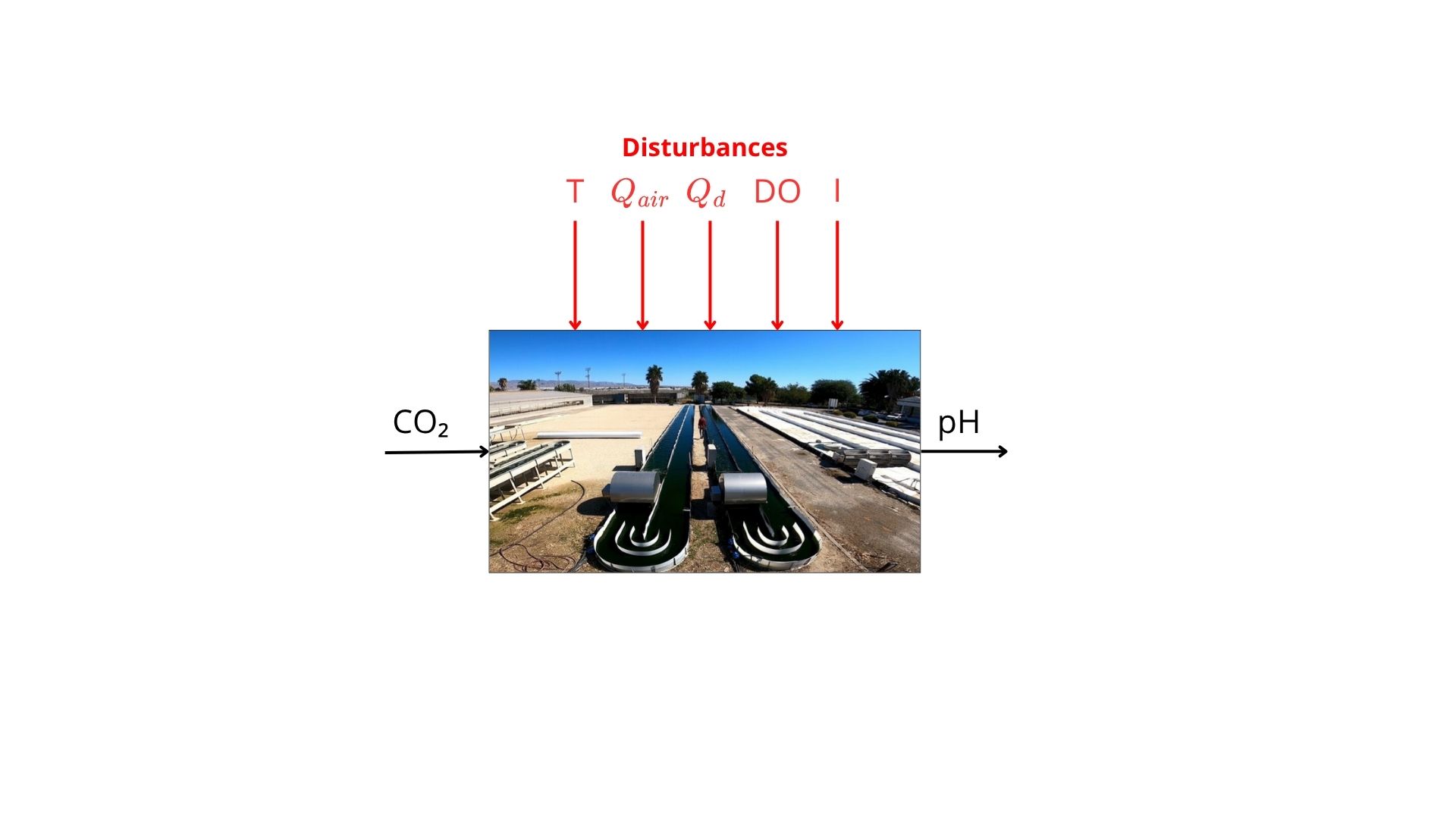}\\
  \caption{External representation of the PBR system.}\label{fig3}
\end{figure}

Nevertheless, accurately capturing pH dynamics requires complex measurements of several biological variables that are either unavailable online or unsuitable from a process control perspective, such as biomass concentration. The lack of these measurements greatly complicates precise modeling and the development of model-based control strategies, thereby justifying the use of techniques such as RL. From the perspective of an external representation of the system, and considering the available measurements in the existing facility, the pH control problem in the PBR can be represented as shown in Fig.~\ref{fig3}.

In this representation, the previously described variables are supplemented with two additional problem disturbances, the air injection flow rate (Q$_{\mathrm{air}}$), which is used to maintain DO at desired levels (disturbing pH), and the dilution flow rate (Q$_{\mathrm{d}}$). The dilution flow is added to the reactor after biomass harvesting or when the culture level drops due to evaporation, and thus does not follow any predefined pattern. These flow rates directly influence mass transfer and concentration balances within the system, affecting key parameters such as pH, thereby playing a crucial role in the overall dynamics and control of the bioprocess.

Thus, the overall control problem consists of maintaining the optimal pH value for the microalgae strain being cultivated by regulating CO$_2$ injection. This control aims to reject the effects of all disturbances impacting pH, as well as to compensate for the inherent dynamics of the photosynthesis process.

\subsection{Reinforcement learning background}

\subsubsection{Reinforcement learning}

\begin{figure}[h]
\centering
  \includegraphics[width=0.7\linewidth, clip, trim=5cm 3cm 4cm 0cm]{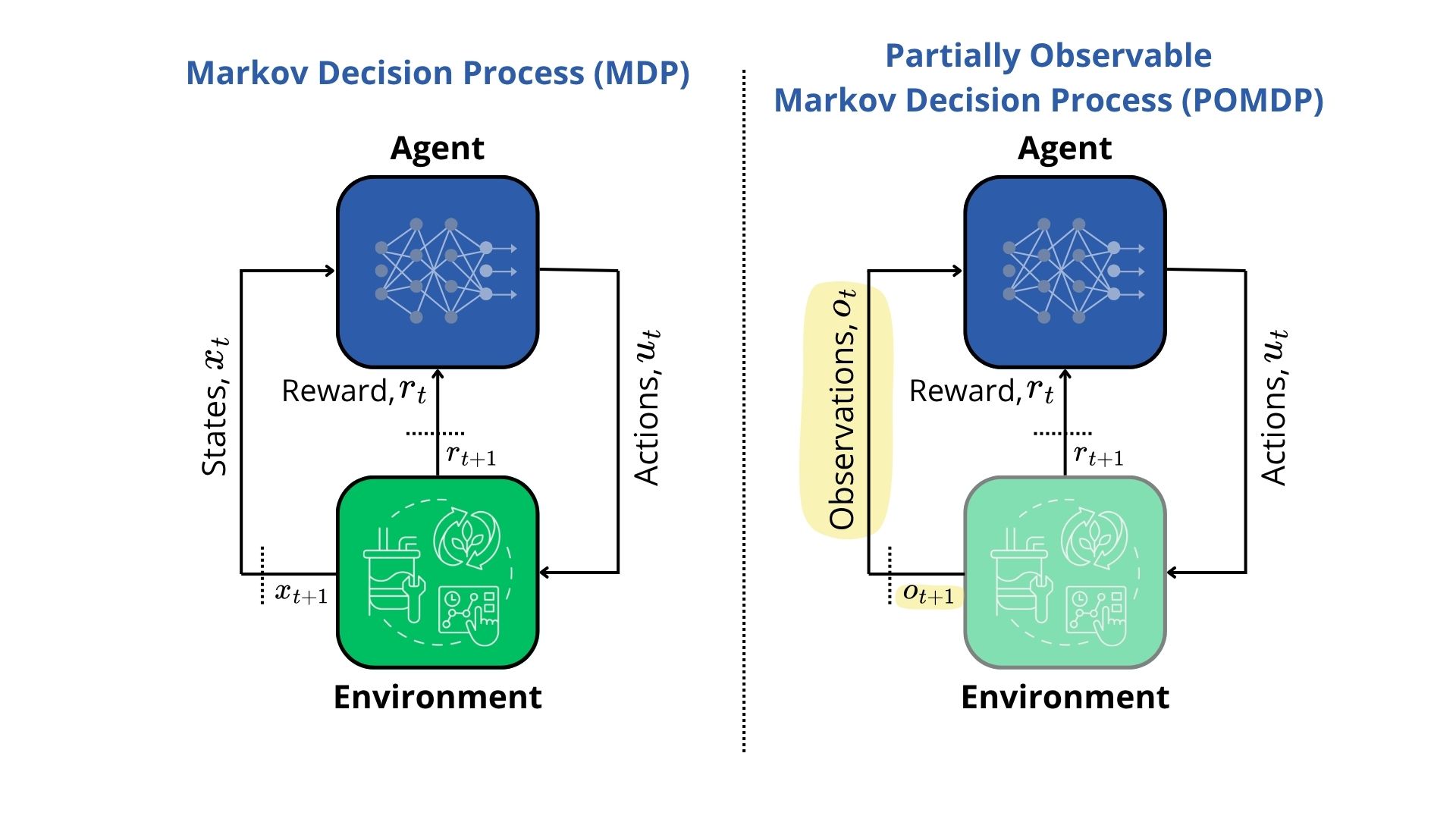}\\
  \caption{Reinforcement learning framework. The left diagram shows the framework for an MDP, and the right one shows the framework for a POMDP.}\label{fig4}
\end{figure}

Most RL algorithms assume that the environment can be modeled as a Markov Decision Process (MDP), formally defined by the quintuple \((\mathcal{X}, \mathcal{U}, P, R, \gamma)\), where \(\mathcal{X}\) is the set of system states, \(\mathcal{U}\) the set of available actions, \(P: \mathcal{X} \times \mathcal{U} \times \mathcal{X} \rightarrow [0,1]\) the state transition probability function, \(R: \mathcal{X} \times \mathcal{U} \rightarrow \mathbb{R}\) the reward function, and \(\gamma \in [0,1]\) the discount factor. In this setting, an agent interacts with the environment over discrete time steps: at each step \(t\), it observes the current state \(\mathbf{x}_t \in \mathcal{X} \subseteq \mathbb{R}^{n_x}\), selects an action \(\mathbf{u}_t \in \mathcal{U} \subseteq \mathbb{R}^{n_u}\) following a policy, \(\pi\), and the environment transitions to a new state \(\mathbf{x}_{t+1}\) accordingly. Fig~\ref{fig4} illustrates this framework.

However, in real-world scenarios—particularly in complex systems such as bioprocesses—full state observability is often unrealistic. Instead, the agent relies on observations \(\mathbf{o}_t \in \mathcal{O} \subseteq \mathbb{R}^{n_o}\), which provide partial information about the underlying true state. Although it does not fundamentally alter the general interaction scheme of RL, partial observability necessitates a modified nomenclature, as depicted in Fig.~\ref{fig4}. This scenario leads to the formulation of a POMDP. From now on, the POMDP formulation will be maintained in this work, as it is the one that defines the PBR system.

To configure the RL agent within this POMDP framework, a commonly adopted approach is the \textit{actor-critic} architecture, which integrates two components: an \textit{actor}, responsible for selecting actions \(\mathbf{u}_t\) based on observations \(\mathbf{o}_t\) via a deterministic policy \( \pi(\mathbf{o}_t; \boldsymbol{\theta}) \); and a \textit{critic}, which evaluates these actions using an action-value function \( Q(\mathbf{o}_t, \mathbf{u}_t; \boldsymbol{\Phi}) \). In this formulation, the parameters \( \boldsymbol{\theta} \) and \( \boldsymbol{\Phi} \) denote the weights of the actor and critic networks, respectively, and are updated during training to optimize both the action-selection policy and its evaluation.

\subsubsection{Deep Deterministic Policy Gradient}

Among the most representative algorithms that implement the \textit{actor-critic} paradigm in continuous action spaces is DDPG \citep{rajasekhar2025exploring}. DDPG specifically relies on neural networks to parameterize both the deterministic policy and the action-value function, making it well-suited for high-dimensional control tasks.

For offline training of an agent following this algorithm, past experiences are stored in a memory structure known as the experience \textit{buffer}, which consists of recorded observations, actions, and rewards at specific sampling times. During training, random mini-batches of size \( M \) are sampled from the buffer to update the \textit{actor} and \textit{critic} models. Specifically, the \textit{critic} is trained by minimizing the following loss function:
\begin{align}
L(\boldsymbol{\Phi}) &= \frac{1}{M} \sum_{i=1}^{M} \left( y_i - Q( \mathbf{o}_i, \mathbf{u}_i; \boldsymbol{\Phi}) \right)^2, \\
y_i &= r_i + \gamma Q_T(\mathbf{o}_{i+1}, \pi_T(\mathbf{o}_{i+1}; \boldsymbol{\theta}_T); \boldsymbol{\Phi}_T). 
\end{align}

\noindent The target value \( y_i \) is computed by summing the immediate reward \( r_i \) and the discounted future return, estimated using decoupled target networks \( Q_T \) and \( \pi_T \). These networks are slowly updated copies of the main networks, used to stabilize learning and prevent oscillations in value estimates \citep{lillicrap2015continuous}.

The \textit{actor}, in turn, is updated by maximizing the expected accumulated reward using the following estimated gradient over a mini-batch of size \( M \):
\begin{equation}
\label{eq3}
\nabla_{\theta} J \approx \frac{1}{M} \sum_{i=1}^{M} \nabla_{\mathbf{u}_i} Q(\mathbf{o}_i, \mathbf{u}_i; \mathbf{\Phi}) \big|_{\mathbf{u}_i =  \pi(\mathbf{o}_i; \boldsymbol{\theta})} \nabla_{\theta} \pi(\mathbf{o}_i; \boldsymbol{\theta}),
\end{equation}
\noindent where the first part of the expression in the sum denotes the gradient of the \textit{critic}'s output with respect to the action generated by the \textit{actor} network, whereas the second corresponds to the gradient of the \textit{actor}'s output with respect to its own parameters.

Note that the updates of the target network parameters are typically performed using a soft update strategy, as follows:
\begin{align}
\nonumber
\boldsymbol{\theta}_T & \leftarrow \tau \boldsymbol{\theta} + (1-\tau) \boldsymbol{\theta}_T, \\ 
\boldsymbol{\Phi}_T & \leftarrow \tau \boldsymbol{\Phi} + (1-\tau) \boldsymbol{\Phi}_T,
 \end{align}
where $\tau$ is the smoothing factor.

The algorithm table detailing the implementation of DDPG is provided in Algorithm~\ref{Algori1}.

\begin{algorithm}[H]
\SetAlgoLined
\KwIn{Offline dataset $\mathcal{D} = \{(\mathbf{o}_j, \mathbf{u}_j, r_j, \mathbf{o}_{j+1})\}_{j=1}^N$, actor network $\pi$, critic network $Q$, target networks $\pi_T$ and $Q_T$, smoothing factor $\tau$, discount factor $\gamma$, mini-batch size $M$}
\textbf{Initialize:} Randomly initialize critic network $Q(\mathbf{o}, \mathbf{u}; \boldsymbol{\Phi})$ and actor $\pi(\mathbf{o}; \boldsymbol{\theta})$ with weights $\boldsymbol{\Phi}$ and $\boldsymbol{\theta}$\; 
Initialize target networks $Q_T$ and $\pi_T$ with weights $\boldsymbol{\Phi}_T \leftarrow \boldsymbol{\Phi}$, $\boldsymbol{\theta}_T \leftarrow \boldsymbol{\theta}$;
 \For{each training iteration}{
  Sample a mini-batch of transitions $\{(\mathbf{o}_i, \mathbf{u}_i, r_i, \mathbf{o}_{i+1})\}_{i=1}^M$ randomly from $\mathcal{D}$\;
  Compute target Q-values for critic update: $y_i = r_i + \gamma Q_T(\mathbf{o}_{i+1}, \pi_T(\mathbf{o}_{i+1}; \boldsymbol{\theta}_T); \boldsymbol{\Phi}_T)$
  
  Update critic by minimizing loss: 
 $L(\boldsymbol{\Phi}) = \frac{1}{M} \sum_{i=1}^{M} \left( y_i - Q(\mathbf{o}_i, \mathbf{u}_i; \boldsymbol{\Phi}) \right)^2 $
 
  Update actor policy using policy gradient: 
$\nabla_{\theta} J \approx \frac{1}{M} \sum_{i=1}^{M} \nabla_{u_i} Q(\mathbf{o}_i, \mathbf{u}_i; \mathbf{\Phi}) \big|_{\mathbf{u}_i =  \pi(\mathbf{o}_i; \boldsymbol{\theta})} \nabla_{\theta} \pi(\mathbf{o}_i; \boldsymbol{\theta})$

  Soft-update target networks with factor $\tau \ll 1$: 
$\boldsymbol{\theta}_T  \leftarrow \tau \boldsymbol{\theta} + (1-\tau) \boldsymbol{\theta}_T, $ 
$\boldsymbol{\Phi}_T \leftarrow \tau \boldsymbol{\Phi} + (1-\tau) \boldsymbol{\Phi}_T,$
 }
\caption{DDPG agent for offline training}
\label{Algori1}
\end{algorithm}

\section{RL-Based control strategy for \textit{raceway} PBR systems}
\label{section3}

This section describes the proposed methodology to apply a DDPG based RL agent to a multi-perturbed and highly nonlinear system, such as the PBR. First, the POMDP formulation is explained in detail, including the resulting control scheme representing the interaction of the agent with the real facility. Then, the methodology to train the agent and to transition to the fine-tuning is presented.

\subsection{POMDP formulation}

\subsubsection{Observation space}

Observation space design in the context of microalgae control presents a significant challenge. As commented, the problem naturally takes the form of a POMDP: many critical variables are not directly observable in real time and must be inferred from other available variables. An effective observation vector therefore has to fuse indirect measurements, domain knowledge, and control‑engineering context so that the RL agent can infer the hidden states and learn policies that remain robust in real operating conditions. The proposed observation space for pH control in PBR systems is organized into three fundamental domains:

\begin{itemize}
    \item \textbf{Variables measured directly from the process:} The first component of the observation vector consists of variables that can be directly measured from the PBR system in real time. According to the schematic shown in Fig.~\ref{fig3}, these include the reactor temperature (T), irradiance (I), dissolved oxygen (DO), dilution flow rate ($Q_d$), air flow rate ($Q_{air}$), and CO$_2$ injection rate. These measurements reflect the physicochemical environment of the culture, the weather conditions, and the current control actions applied to the system. As such, they form the backbone of the observable state and provide the RL agent with essential, real-time information to infer the underlying process dynamics and make informed control decisions.

    \item \textbf{Temporal information:} The second component of the observation vector incorporates temporal information, which is essential due to the periodic nature of PBR operation. Microalgae cultivation is tightly coupled to the day–night cycle, as photosynthesis depends on light availability and follows a predictable rhythm. Thus, including time-related features—such as time of day—enables the RL agent to align its decisions with the photosynthetic cycle.

    \item \textbf{Control variables:} The third component of the observation vector is related to control variables. In addition to including current actuator values, this component incorporates the control error (i.e., the difference between the current pH and its setpoint) as well as the integral of the error over time. These terms are standard in classical control theory, particularly in PID-based strategies, as they capture both the instantaneous deviation and the accumulated offset from the desired value. Including them in the observation space provides the RL agent with a sense of performance history and helps guide the learning of corrective actions that not only stabilize the system but also minimize long-term deviations, making the policy more effective in a control regulation problem context.
\end{itemize}

It is worth noting that some authors in the literature augment the observation space with historical measurements to address the issue of partial observability. However, in this case—given that it is a regulation control problem—this historical information is captured through control-related variables, particularly the integral of the error. This approach simplifies the implementation of the DDPG agent.

\subsubsection{Action space}

The action space in this formulation is defined by the CO$_2$ injection rate, as this is the primary actuator available for pH regulation in the PBR system. Injecting CO$_2$ into the medium increases the concentration of dissolved carbon dioxide, which shifts the carbonate equilibrium and leads to a reduction in pH. 

It should be noted that the action space is constrained between a minimum and maximum CO$_2$ injection rate, i.e., for each sampling time $t$,
$\mathrm{CO}_2^{min} \leq \mathrm{CO}_{2,t}  \leq \mathrm{CO}_2^{max}$, reflecting the physical limitations of a real actuator used in the PBR system. This limitation can be directly integrated into the agent’s action space without introducing additional complexity. However, it is important to consider that actuator saturation may lead to integral windup in the error integration term included in the observation space. This phenomenon occurs when the integral component continues to accumulate error even though the actuator is out of its limits, potentially causing instability or delayed recovery once the system re-enters the controllable range. To mitigate this issue, an appropriate anti-windup mechanism based in clipping the integral term when is implemented in this study \citep{caparroz2025hybrid}. 

\subsubsection{Reward function}

Since the task is focused on control purposes, the most straightforward approach is to use a reward function that directly depends on the error, such as the quadratic error, defined as:
\begin{equation}
   r_t = - (e_t)^2 
\end{equation}
\noindent where $e_t = pH_{\text{SP}} - pH_t$ represents the control error at sampling time $t$ between the desired set-point $pH_{\text{SP}}$ and the current pH value, $pH_t$. However, as the agent will be trained with data from a PID controller, where the error is expected to be closed to zero, this type of function can pose problems when calculating the gradients for training the agent in the DDPG algorithm (see Eq.~\ref{eq3}). For this reason, a logarithmic function of the error is proposed instead. This function is given by: 
\begin{equation}
   r_t = - log(e_t^2 + \epsilon), 
\end{equation}
\noindent where $\epsilon$ is a parameter that adjust the maximum of the function (i.e., the maximum value that is reached when the error is zero).

\begin{figure}[h]
\centering
  \includegraphics[width=0.8\linewidth, clip, trim=0cm 0cm 0cm 0cm]{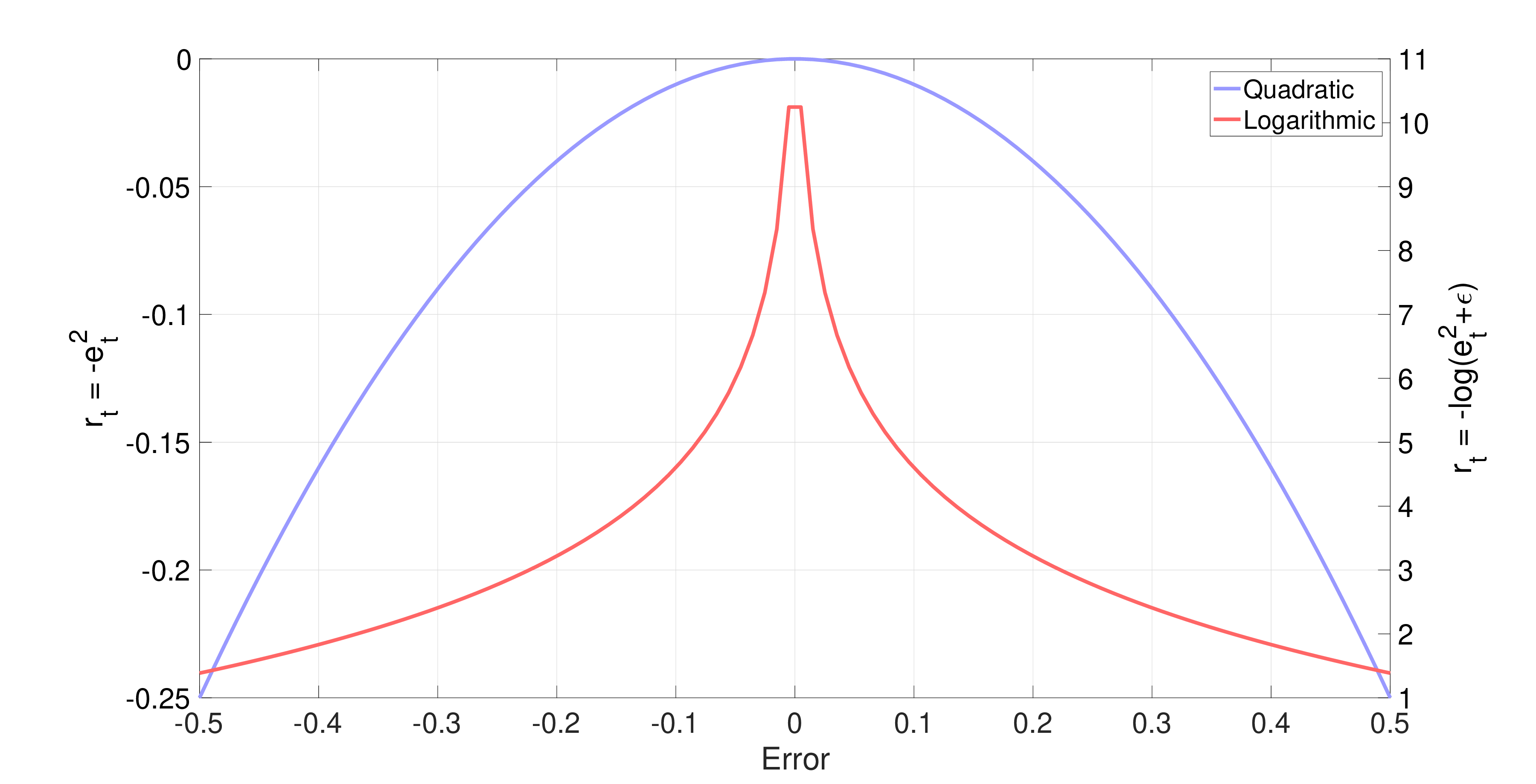}\\
  \caption{Comparison of rewards function with respect to the error. The logarithmic function is set with $\epsilon=10^6$.}\label{fig5}
\end{figure}

Figure~\ref{fig5} illustrates the differences in reward values between the proposed logarithmic function and the quadratic function across varying error magnitudes. The logarithmic reward function reaches its maximum (i.e., the value achieved with an $\epsilon$ of $10^6$) when the error is near zero. As the error subsequently increases, the function penalizes these deviations but smooths out large errors, a characteristic crucial in contexts like open PBRs. It is important to note that open PBR systems are exposed to environmental conditions, unmeasurable disturbances, and routine operational considerations, such as sensor's calibration, which can cause large transient errors that, in turn, can destabilize the control loop. Conversely, it can be observed that the quadratic error function maintains values very close to zero with small errors, making it difficult to observe changes in the function's gradient, and on the other hand, it quadratically penalizes large errors.

\subsubsection{Resulting block diagram}
\begin{figure}[!h]
\centering
  \includegraphics[width=0.85\linewidth, clip, trim=0cm 6cm 7cm 2cm]{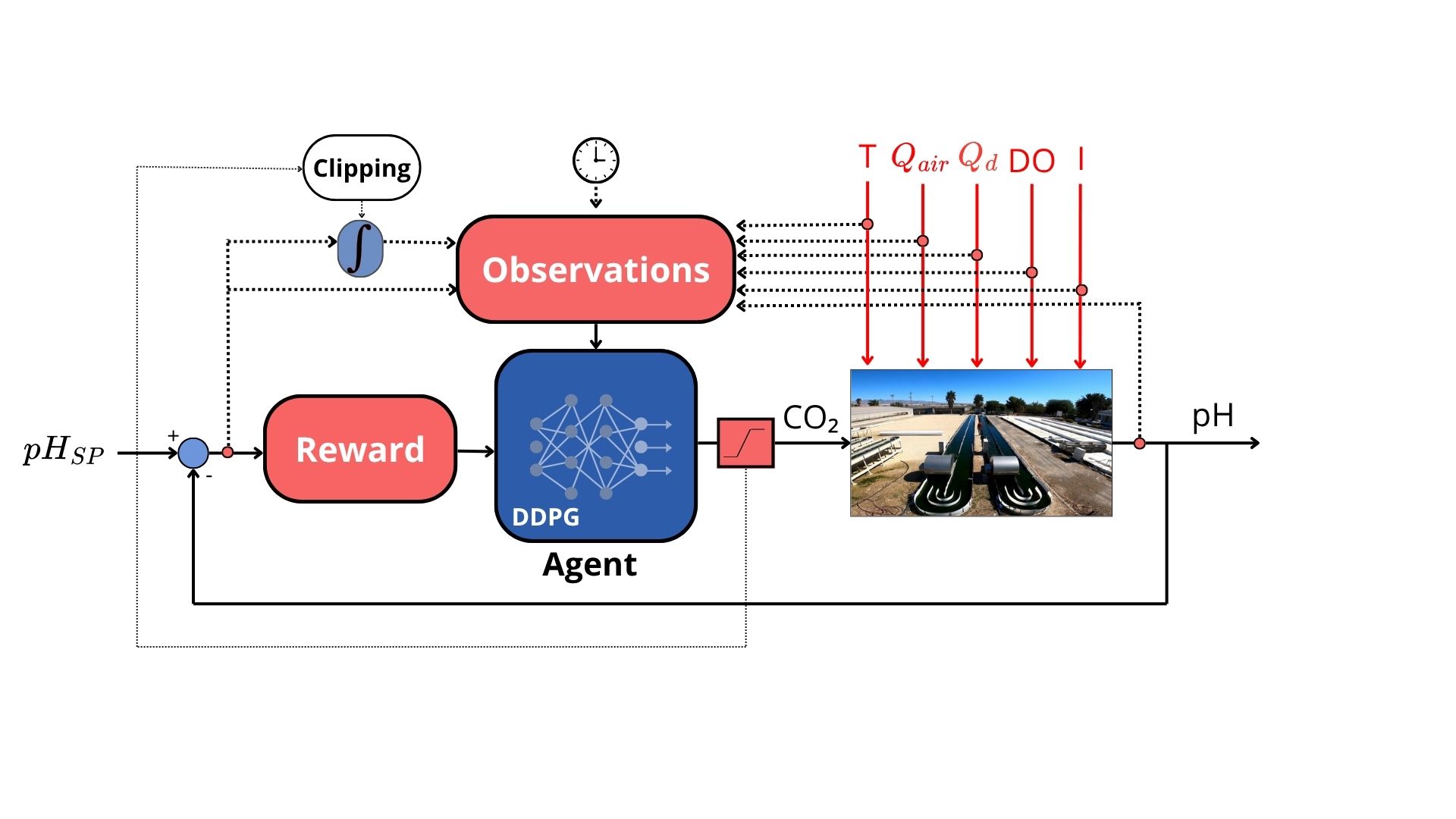}\\
  \caption{Proposed control structure.}\label{fig6}
\end{figure}

Taking into account the different components described in the previous subsections, the resulting block diagram representing the interaction between the agent and the real facility is presented in Fig.~\ref{fig6}. As depicted in the figure, the agent operates within a classical feedback control scheme, with its behavior guided by the reward function. A key feature, however, is that the observation space—fed by direct measurements of process variables—effectively acts as a feedforward control structure, enabling the rejection of measurable disturbances \citep{guzman2024feedforward}. It is also worth noting that a clipping structure has been added to the error integral to prevent windup issues when the system enters actuator saturation.

\subsection{RL algorithm training}

The methodology proposed in this work combines the offline training of the RL agent with subsequent online fine-tuning, and it is fully described in Algorithm~\ref{algorit2}. The offline component enables the agent to learn an initial control policy from historical data generated by an expert system, specifically a PID controller. The online fine-tuning phase, on the other hand, serves a dual objective: i) to adapt the agent's policy to the changing dynamics inherent in the PBR system, and ii) to progressively learn from the effects of disturbances on the system. This continuous adaptation aims to significantly enhance the performance beyond that of simple PID-type controllers, from which the agent initially acquired its knowledge.

It is important to note that online fine-tuning of a DDPG agent may lead to overfitting or destabilization, particularly due to excessive updates to the actor and critic networks. To mitigate this, the number of training epochs is limited during this phase, thereby constraining the frequency of gradient-based updates. This enables the agent to gradually adapt to changing conditions while maintaining policy stability and overall robustness.

Furthermore, during experience replay in the fine-tuning phase (see 2 in Step 3 of Algorithm~\ref{algorit2}), the replay buffer is initially populated with historical data, which is then progressively replaced by new experiences collected during operation. This approach aims to dynamically update the agent’s policy using recent data that reflects operating conditions specific to the time of year and other dynamic characteristics of the bioprocess itself.
\\

\begin{algorithm}[H]
\label{algorit2}
\caption{Development of the DDPG agent in an offline environment with real time fine-tuning}

\textbf{Input:} Historical dataset $\mathcal{D}$ of the PBR under diverse operating conditions.

\textbf{Output:} Robust and optimized control policy for pH regulation in the PBR.

\textbf{Step 1:} Preparation of the historical dataset $\mathcal{D}$:
\begin{enumerate}
\setlength{\itemsep}{0pt}
\item Collect historical operating data from the PBR using the expert system.
\item Preprocess the collected data and organize it into observation-action-reward tuples, that is, $\mathcal{D} = \{(\mathbf{o}_j, \mathbf{u}_j, r_j, \mathbf{o}_{j+1})\}_{j=1}^N$.
\end{enumerate}

\vspace{1pt}

\textbf{Step 2:} Offline training of the DDPG agent:
\begin{enumerate}
\setlength{\itemsep}{0pt}
\item Randomly initialize the neural network parameters of the DDPG agent.
\item Apply the training process detailed in Algorithm~\ref{Algori1}.
\end{enumerate}

\vspace{1pt}

\textbf{Step 3:} Online fine-tuning of the DDPG agent:
\begin{enumerate}
\setlength{\itemsep}{0pt}
\item Initialize the experience buffer using the historical dataset $\mathcal{D}$.
\item Continuously collect new online experiences during bioprocess operation at each sampling time $t$, that is, $\mathcal{D}^{*}_t = (\mathbf{o}_t, \mathbf{u}_t, r_t, \mathbf{o}_{t+1})$. This will form the data set of new experiences: $\mathcal{D}^{New} = \{(\mathbf{o}_j, \mathbf{u}_j, r_j, \mathbf{o}_{j+1})\}_{j=1}^{N^{New}}$
\item Dynamically update the experience buffer, retaining the most recent experiences.
\item Perform fine-tuning of the DDPG agent using the training process described in Algorithm~\ref{Algori1}\footnote{The training for fine-tuning follows Algorithm 1, but with a limited number of iterations or epochs to avoid overfitting issues or destabilization of the agent.}. 
\end{enumerate}
\end{algorithm}
\vspace{0.5cm}

\section{Results}
\label{section4}
In this section, the computational implementation of the proposed RL-based control scheme is first described. Then, a simulation study is presented to highlight the benefits of the proposed approach compared to a PID controller, the expert system from which the RL agent was trained, and an RL agent without fine-tuning. Finally, the results obtained from the real facility during eight days of operation are presented.

\subsection{Computational implementation}

\begin{figure}[h]
\centering
  \includegraphics[width=0.8\linewidth, clip, trim=12cm 6cm 7cm 2cm]{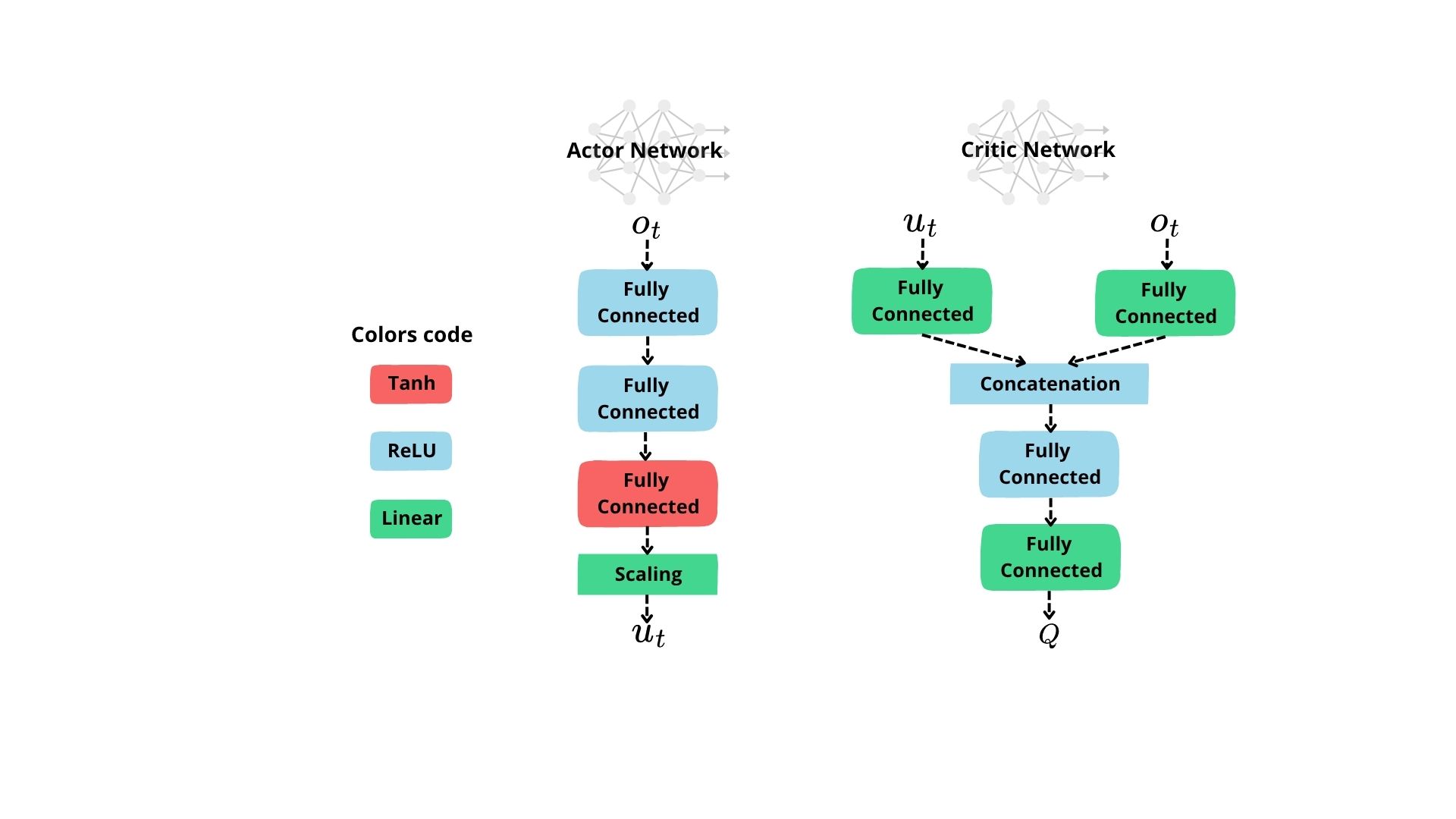}\\
  \caption{Actor and critic structure.}\label{fig7}
\end{figure}

With respect to the agent's architecture, the actor and critic networks were established as represented in Fig.~\ref{fig7}. The actor network consists of a recurrent neural network with three fully connected layers. The first two use a ReLU activation function, while the third is followed by a Tanh activation function to ensure that the output actions remain within a bounded range. Finally, a scaling layer maps the Tanh output to the environment-specific action space, which, in this case, was between 0 and 10 L/min according to the actuator physical limits. On the other hand, the critic network consists of a recurrent neural network as well, and it employs a dual-input architecture designed to estimate the Q-value for a given observation-action pair. It employs two feature input layers: one for the observation and one for the action. These inputs are processed separately through fully connected layers and then merged using a concatenation layer. The merged representation is passed through additional fully connected layers with ReLU activations, culminating in a final fully connected layer that outputs the scalar Q-value estimation. It should be noted that each layer, both in the actor and critic, contains 256 neurons. 

\begin{figure}[h]
\centering
  \includegraphics[width=0.78\linewidth, clip, trim=9cm 0cm 9cm 0cm]{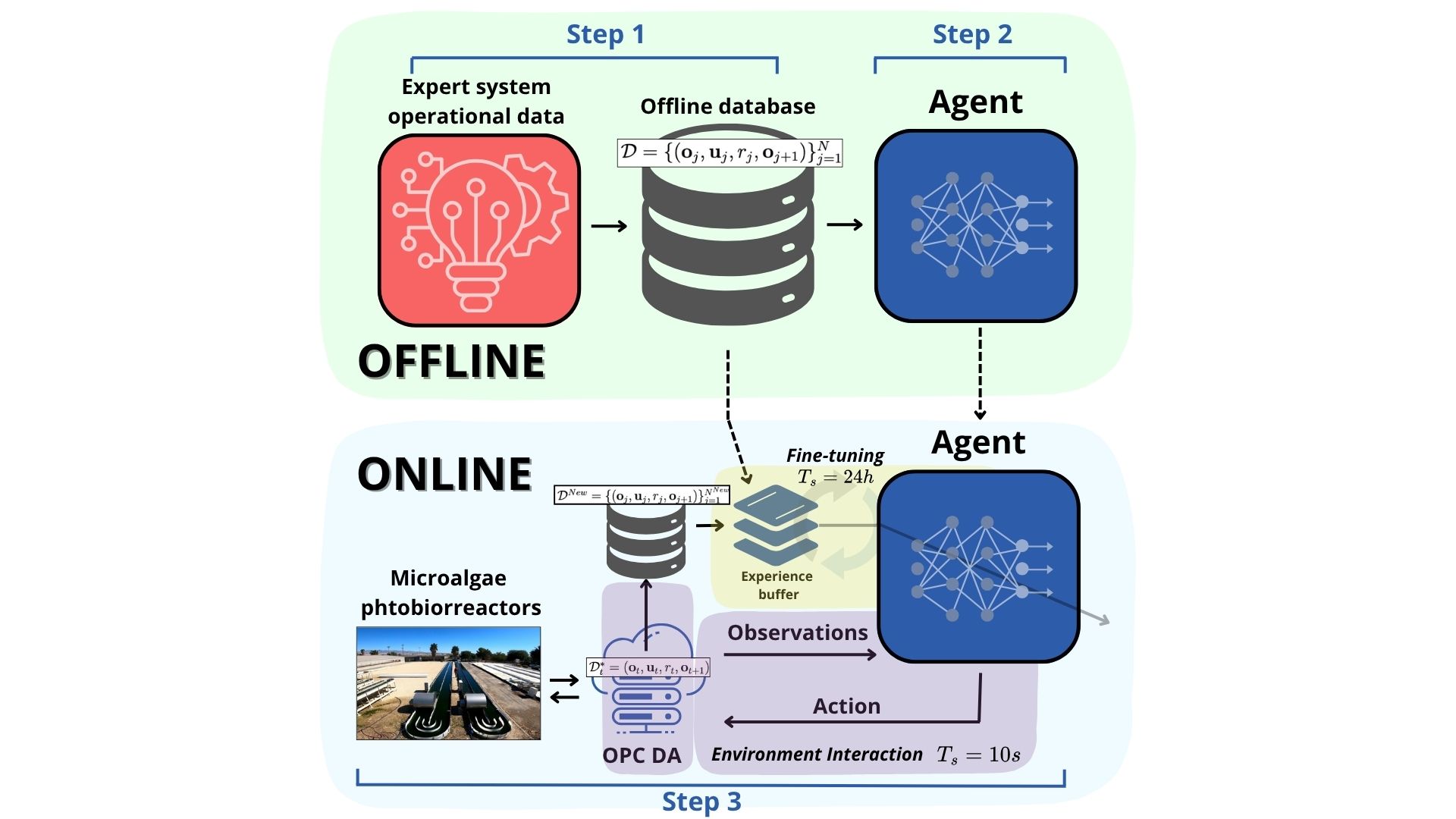}\\
  \caption{Flow-chart representing the implementation of the proposed methodology. In the online phase, the red area represents the agent–environment interaction, which occurs with  $T_s=$10 s, while the yellow area represents the fine-tuning process, which is performed once every 24 hours.}\label{fig8}
\end{figure}

Regarding the parameters of the DDPG algorithm, the agent was configured using the Adam optimization method~\citep{kingma2014adam}, with learning rates of $10^{-4}$ for the critic and $10^{-5}$ for the actor. Moreover, the discount factor ($\gamma$) was set to $0.9$, the smoothing target factor ($\tau$) at 0.01, and the sampling time was set to $10$~s, in accordance with that of the actual facility. The mini batch size ($M$) was set at 64.

The entire implementation was carried out in MATLAB 2023a~\citep{MatlabOTB}. To deploy Algorithm~\ref{algorit2}, the scheme shown in Fig.~\ref{fig8} was used. Historical data from the expert system were obtained using a simple PID controller configured in the ideal form without the derivative term, with parameters \(K_p = -32\) [L/min] and \(T_i = 1200\) [s\(^{-1}\)] \citep{caparroz2023control}. This corresponds to Step 1 of Algorithm~\ref{algorit2}. Using these data, offline training was then conducted over 4000 epochs (Step 2 of Algorithm~\ref{algorit2}). For online deployment and fine-tuning (Step 3), communication with the real facility was established through an industrial protocol, specifically via an OPC DA server. This protocol enabled the retrieval of observations from the real PBR system and the transmission of actions computed by the agent, based on the established sampling time of $T_s = 10$~s. During operation, data were also stored continuously. At the end of each day (23:59~h), fine-tuning was performed, with training limited to 50 epochs in order to prevent potential robustness issues in the agent. As a result, the fine-tuning procedure effectively operated with a 24-hour update cycle.

Finally, it should be mentioned that several well-established operational aspects of the plant were also taken into account in the implementation. First, the pH setpoint was fixed at 8, which is optimal for the type of microalgae strain used. The controller was activated when solar radiation exceeded 100~W/m$^2$ and the actual pH of the system was above the desired reference. It was then deactivated when radiation fell below the defined threshold. Air injections were maintained on demand using a on/off controller integrated into the system’s PLCs, to keep the DO level below a specified limit. This controller was not part of the designed control architecture, as it operated a discontinuous actuator.
However, its effect was indirectly captured in the agent’s observation space through the measurements of DO and the recorded air injection events.

\subsection{Comparative study in a representative simulation environment}

To highlight the advantages of the proposed methodology, a comparative simulation study was conducted using three control strategies: 
\begin{enumerate}
    \item A control strategy based on a nominal PID controller (denoted as PID), with the aim of comparing the agent’s performance with that of the expert system used during training.
    \item A control strategy using an offline-trained DDPG RL agent without online fine-tuning (denoted as RL).
    \item A strategy that incorporated the proposed methodology with daily fine-tuning of the agent's parameters at the end of each day (referred to as RL-FT).
\end{enumerate}

For a fair comparison, this study was carried out using a representative model of the system, validated for the PBR facilities available at the IFAPA research center. The model was fed with meteorological data collected from the actual facility. To increase the difficulty and robustness of the evaluation, the real data used in the study correspond to different time periods: training data were taken from April, while testing data correspond to November 2023. Please note that the used model was described in detail in~\citep{nordio2024abaco, sanchez2021modeling} and is available at~\citep{rodriguez_miranda_2025_15579693}.

\begin{figure}[h]
\centering
  \includegraphics[width=1\linewidth, clip, trim=2cm 1cm 1cm 0cm]{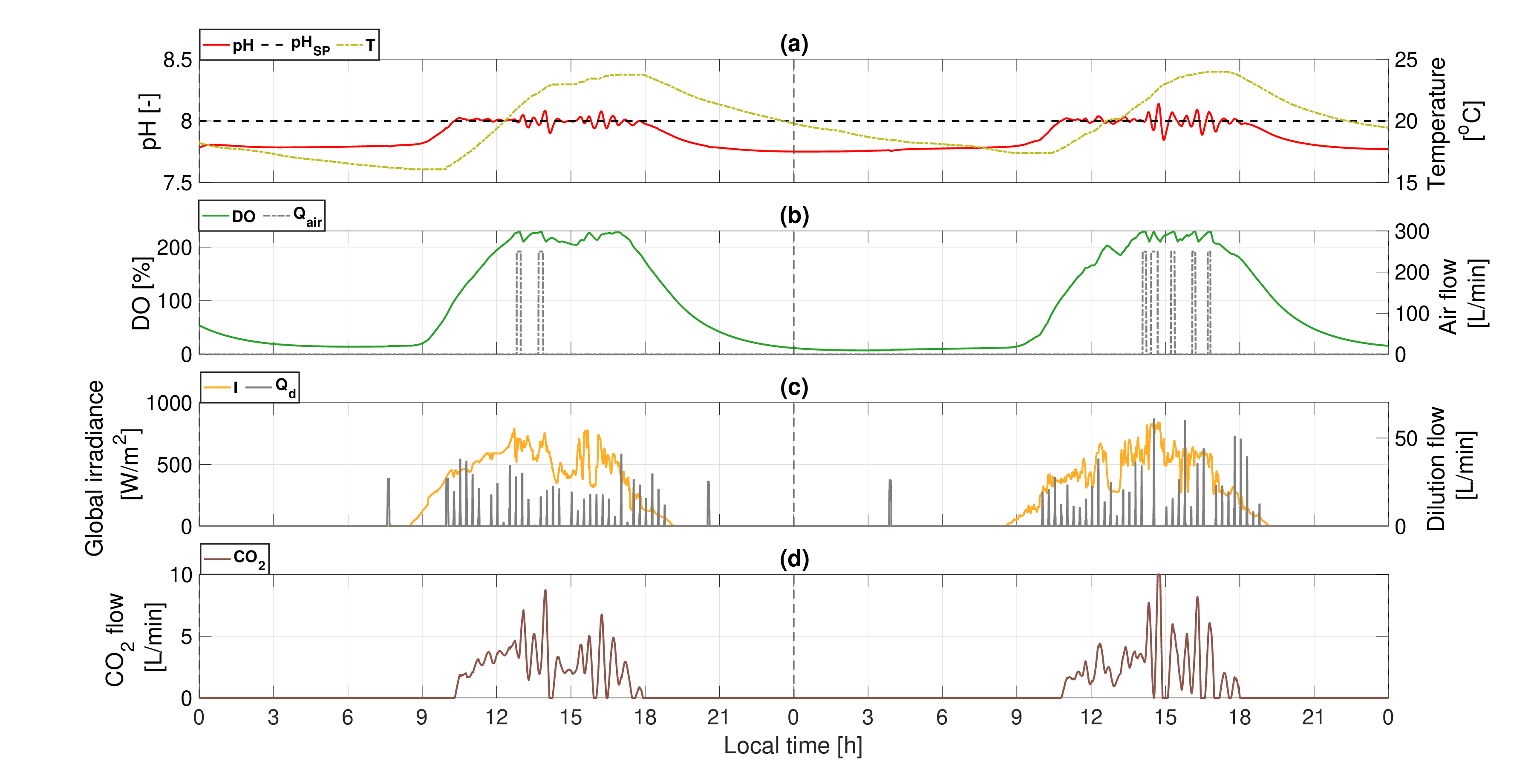}\\
  \caption{Operation of the simulated PBR system using the PID controller. (a) pH in the PBR system (pH), pH reference (pH$_{\mathrm{SP}}$), and PBR system temperature (T); (b) Dissolved oxygen (DO) and air flow injections (Q$_{\mathrm{air}}$); (c) Global irradiance (I), and dilution flow injections (Q$_\mathrm{d}$); (d) CO$_2$ flow (CO$_2$). }\label{fig9}
\end{figure}

Figure~\ref{fig9} presents the training data, i.e., the data obtained with the PID-expert system. In this case, two full days of operation were used for the training phase. As can be observed, during these two days there were variations in the main disturbances affecting the system, such as DO and air injection (see Fig.~\ref{fig9}-(b)), or solar irradiance and dilution flow rate (see Fig.~\ref{fig9}-(c)). To counteract the effects of disturbances and maintain the desired setpoint, the PID controller provided the CO$_2$ injection signal shown in the Fig.~\ref{fig9}-(d).

\begin{figure}[!h]
\centering
  \includegraphics[width=1\linewidth, clip, trim=2cm 1cm 1cm 0cm]{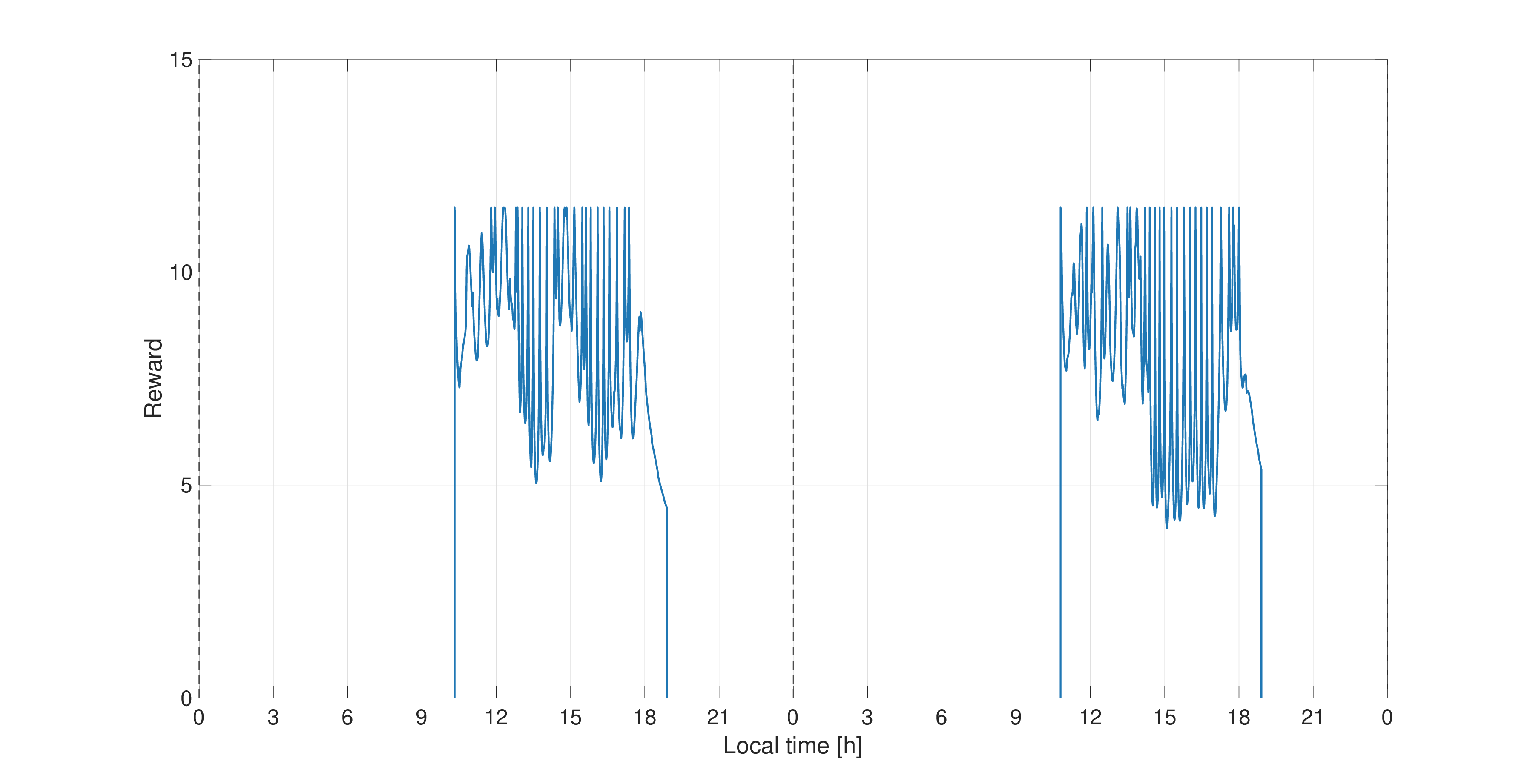}\\
  \caption{Values of the reward function during the PID control of the simulated PBR system.}\label{fig10}
\end{figure}

Moreover, Fig.~\ref{fig10} shows the values of the reward function during operation with the PID controller, which were later used for training the RL agent. Note that the moments when the value was zero correspond to times when the controller was not active, due to the activation conditions mentioned earlier. These instances were not taken into account during the training phase.

\begin{figure}[!h]
\centering
  \includegraphics[width=1\linewidth, clip, trim=2cm 1cm 1cm 0cm]{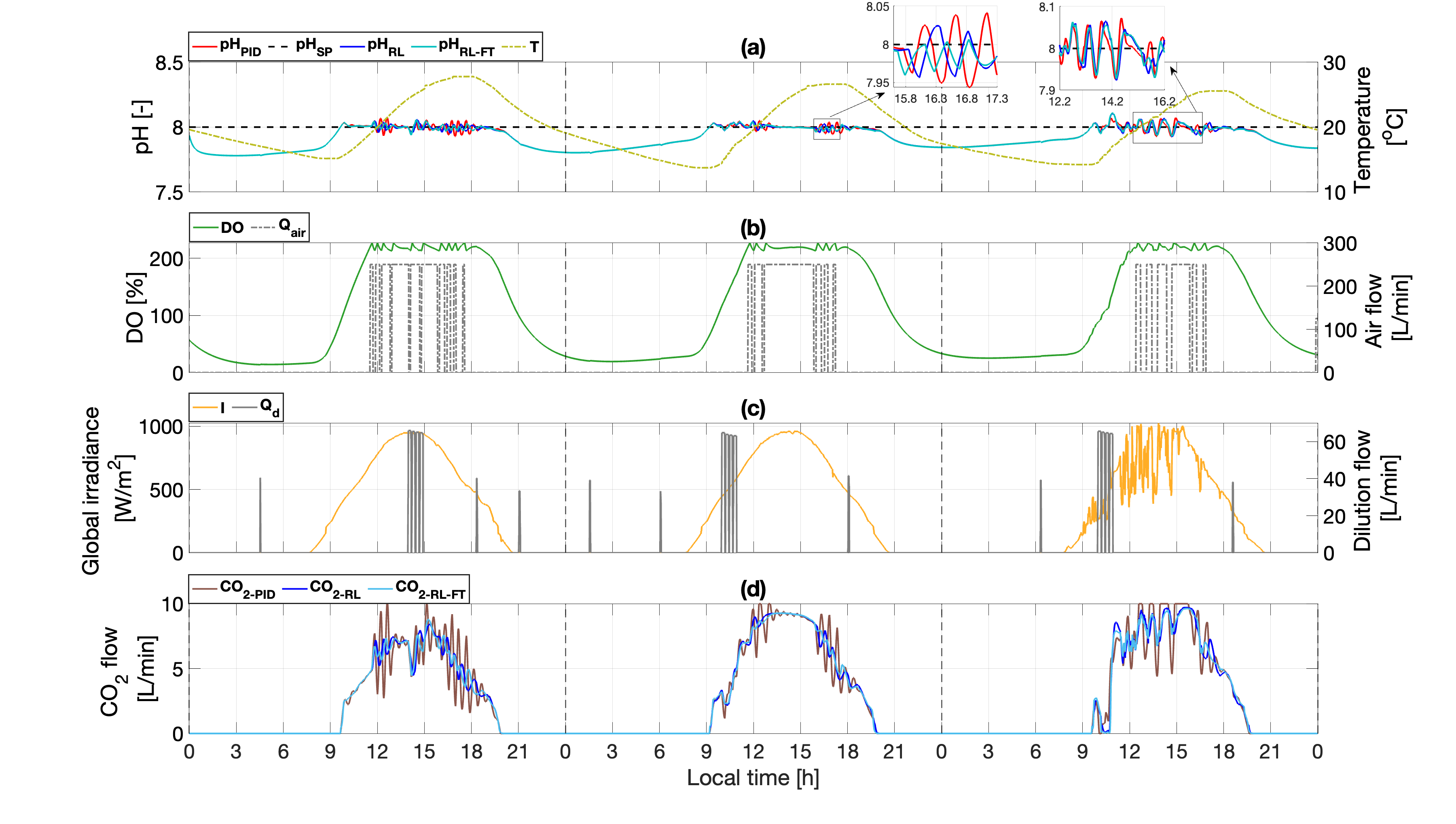}\\
  \caption{Results obtained from the simulated system using the three controllers. (a) pH in the PBR system provided by the PID controller (pH$_\mathrm{PID}$), the RL agent without fine-tuning (pH$_\mathrm{RL}$), and the RL agent with fine-tuning (pH$_\mathrm{RL-FT}$), pH reference (pH$_{\mathrm{SP}}$), and PBR system temperature (T); (b) Dissolved oxygen (DO) and air flow injections (Q$_{\mathrm{air}}$); (c) Global irradiance (I), and dilution flow injections (Q$_\mathrm{d}$); (d) CO$_2$ flow provided by the PID controller (CO$_\mathrm{2-PID}$), the RL agent without fine-tuning (CO$_\mathrm{2-RL}$), and the RL agent with fine-tuning (CO$_\mathrm{2-RL-FT}$).}\label{fig11}
\end{figure}

The results obtained with the three controllers are presented in Fig.~\ref{fig11}. Note that the variables acting as disturbances are the same for all three controllers; therefore, only the pH variable and the CO$_2$ injection signal are shown for the three cases. In this test, data from a different time of year were used, which altered the system dynamics with respect to the ones observed in Fig.~\ref{fig10}. Additionally, the dilution flow injection  and the irradiance profiles (see Fig.~\ref{fig11}-(c)) were different. This resulted in the PID controller exhibiting behavior that differed significantly from that observed during the training days; as can be seen, there were more oscillations in the control signal (see Fig.~\ref{fig11}-(d)). In contrast, for the RL-based strategies, the behavior was less oscillatory. This is not because the RL agent behaves like a filtered version of a PID controller, but rather because it has learned the basic dynamic behavior of the system during the offline training phase, thanks to the proposed observation space. As a result, the RL agent rejects disturbances intrinsically, whereas the PID controller does so solely through feedback.

The effect of fine-tuning is especially noticeable from the second day of operation onward. The zoomed-in views presented in the Fig.~\ref{fig11}-(a) show how the fine-tuned agent improved disturbance rejection. This was due to the agent gradually learning the effect of the disturbances and the new dynamics during the fine-tuning process. This is a valuable result that demonstrates how fine-tuning enables the transition from a fixed policy trained offline to an adaptive policy suited to new operating conditions.

To quantitatively demonstrate the improvements of the proposed methodology, Tab.~\ref{tb:Rangos} shows the Integral of Absolute Error (IAE) and Cumulative Control Effort (CCE) indices for each of the controllers. The IAE measures the accumulated absolute difference between the desired setpoint and the actual process variable, reflecting overall control accuracy. The CCE index quantifies the total amount of control action exerted by the controller during operation. These indices were calculated over the three days of operation,  only for the periods when the controllers were active, as follows: 

\begin{equation}
\text{IAE} = \int_0^T \left| e_t \right| \, dt, \quad \text{CCE} = \int_0^T \left| \Delta u_t \right| \, dt.
\end{equation}

\begin{table}[h]
    \centering
    \caption{Comparison of control metrics}
    \label{tb:Rangos}
    \begin{tabular}{ccc}
        \hline
        \textbf{Controller} & \textbf{IAE} & \textbf{CCE} \\
        \hline
        PID & 2339.1 & 302.43 \\
        RL & 2276.9 & 149.53 \\
        RL-FT & 2162.0 & 140.30 \\
        \hline
    \end{tabular}
\end{table}

The quantitative comparison in Table~\ref{tb:Rangos} demonstrates that RL-FT improved performance in terms of IAE by approximately 8\% compared to the PID strategy and by around 5\% relative to standard RL, within just three days of operation. Furthermore, it also achieved a significant reduction in control effort, attributable to its intrinsically anticipatory control actions and its ability to adapt to new operating conditions. Specifically, the control effort was approximately 54\% lower than that of the PID strategy and 7\% lower than that of the RL agent. This improvement is particularly relevant in process control, where changes in the control signal are typically associated with operating costs—especially in this case, as they are directly linked to pure CO$_2$ injections.

\subsection{Experimental results on the real system}

\begin{figure}[h]
\centering
  \includegraphics[width=1\linewidth, clip, trim=2cm 0cm 1cm 0cm]{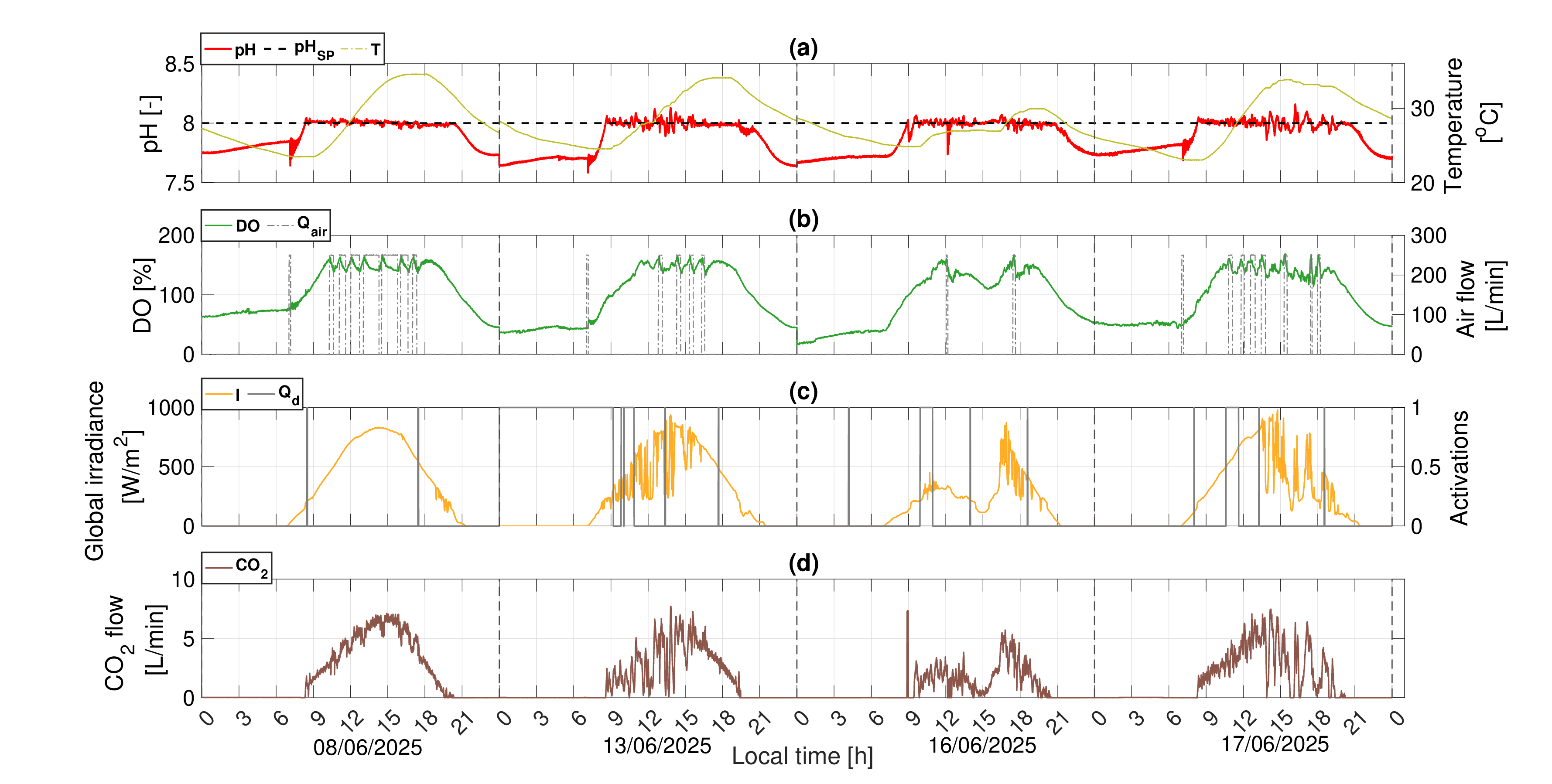}\\
  \caption{Operation of the real PBR system using the PID controller. (a) pH in the PBR system (pH), pH reference (pH$_{\mathrm{SP}}$), and PBR system temperature (T); (b) Dissolved oxygen (DO) and air flow injections (Q$_{\mathrm{air}}$); (c) Global irradiance (I), and dilution flow injections (Q$_\mathrm{d}$); (d) CO$_2$ flow (CO$_2$).}\label{fig12}
\end{figure}

After demonstrating its performance in simulation, the proposed methodology was implemented on the real PBR system during June 2025. First, data from four days of operation using a PID controller in the actual reactor were collected, as shown in Fig.~\ref{fig12}. These four days were not consecutive; instead, data were gathered from days with different operating conditions. For example, the first day corresponded to a Sunday, when plant technicians did not perform harvesting operations, and therefore the dilution flow rate remained relatively stable, with only minor flow injections due to evaporation. The remaining days corresponded to weekdays, during which harvesting was carried out, and the reactor was subject to more variable conditions in terms of dilution flow rate. Moreover, there were also fluctuations in solar radiation. It should be noted that the minimal presence of disturbances on the first day allowed the PID controller to perform relatively well. However, as disturbances arose in the following days, its performance noticeably declined. The maximum deviation with respect to the reference was 0.27~[-], and it was observed in the third day around 12:00~h (see Fig.~\ref{fig12}-(a)), after dilution rate and air flow injections (see Figs.~\ref{fig12}-(b) and (c)).

\begin{figure}[h]
\centering
  \includegraphics[width=1\linewidth, clip, trim=2cm 1cm 1cm 0cm]{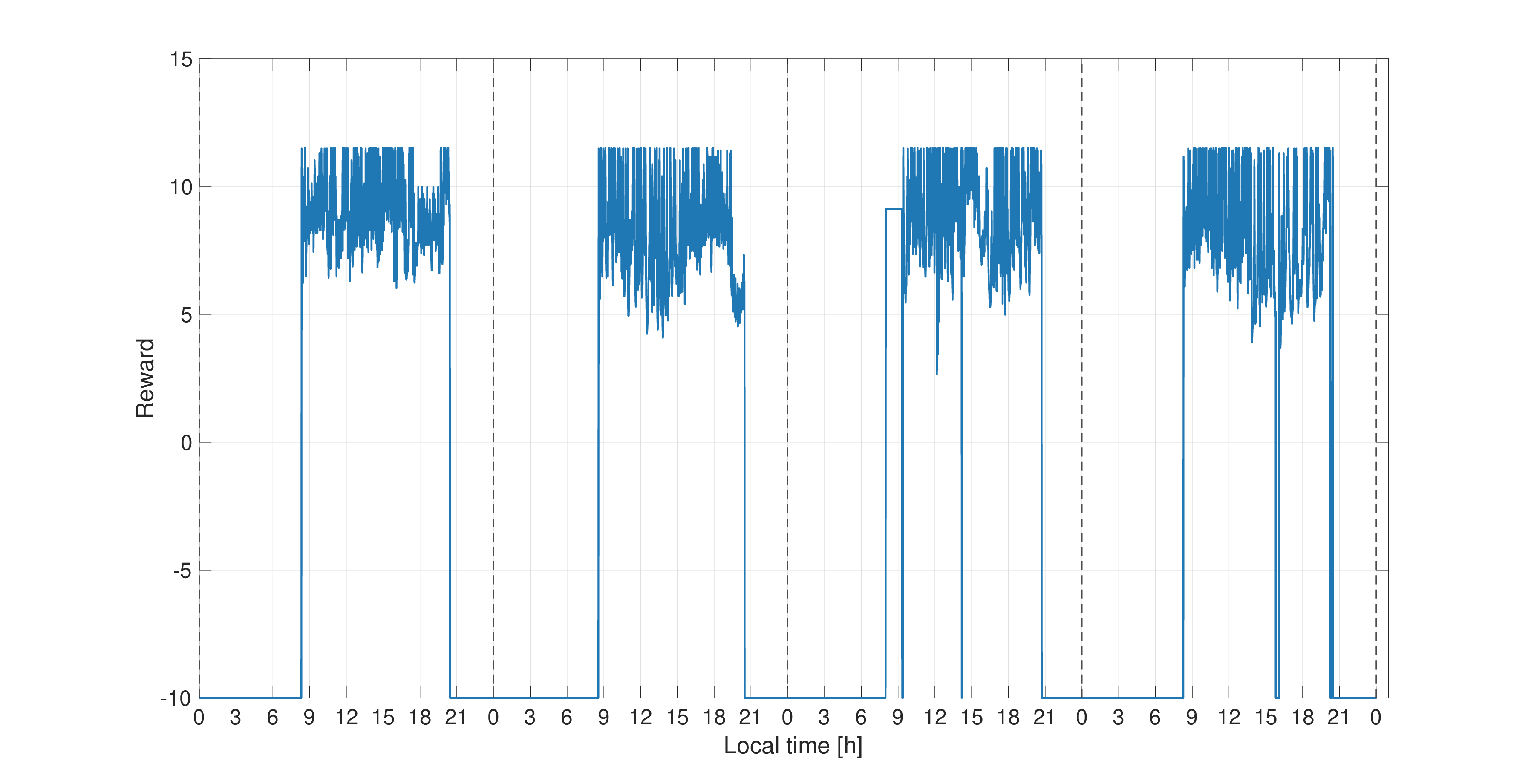}\\
  \caption{Values of the reward function during the PID control of the real PBR system.}\label{fig13}
\end{figure}

Fig.~\ref{fig13} shows the values of the reward function obtained during the PID operation, which were later used for training the agent. As in the simulation, periods when the controller was not operating were directly set to zero and were not considered during training. Additionally, the main characteristics of the reward function used can be observed in this case. For example, on day three, as mentioned, there were several abrupt pH changes due to dilution flow injections (see Fig.\ref{fig12}-(a)), which caused large transient error values. As shown, the reward function penalized these changes with values close to zero, smoothing their impact and avoiding large scale jumps. Furthermore, errors close to zero were amplified—as in the simulation—within a range of approximately 6 to 11, which helps prevent issues in gradient computation during the RL agent training phase.

\begin{landscape}
\begin{figure}[!h]
\centering
  \includegraphics[width=1.1\linewidth, clip, trim=2cm 1cm 1cm 0cm]{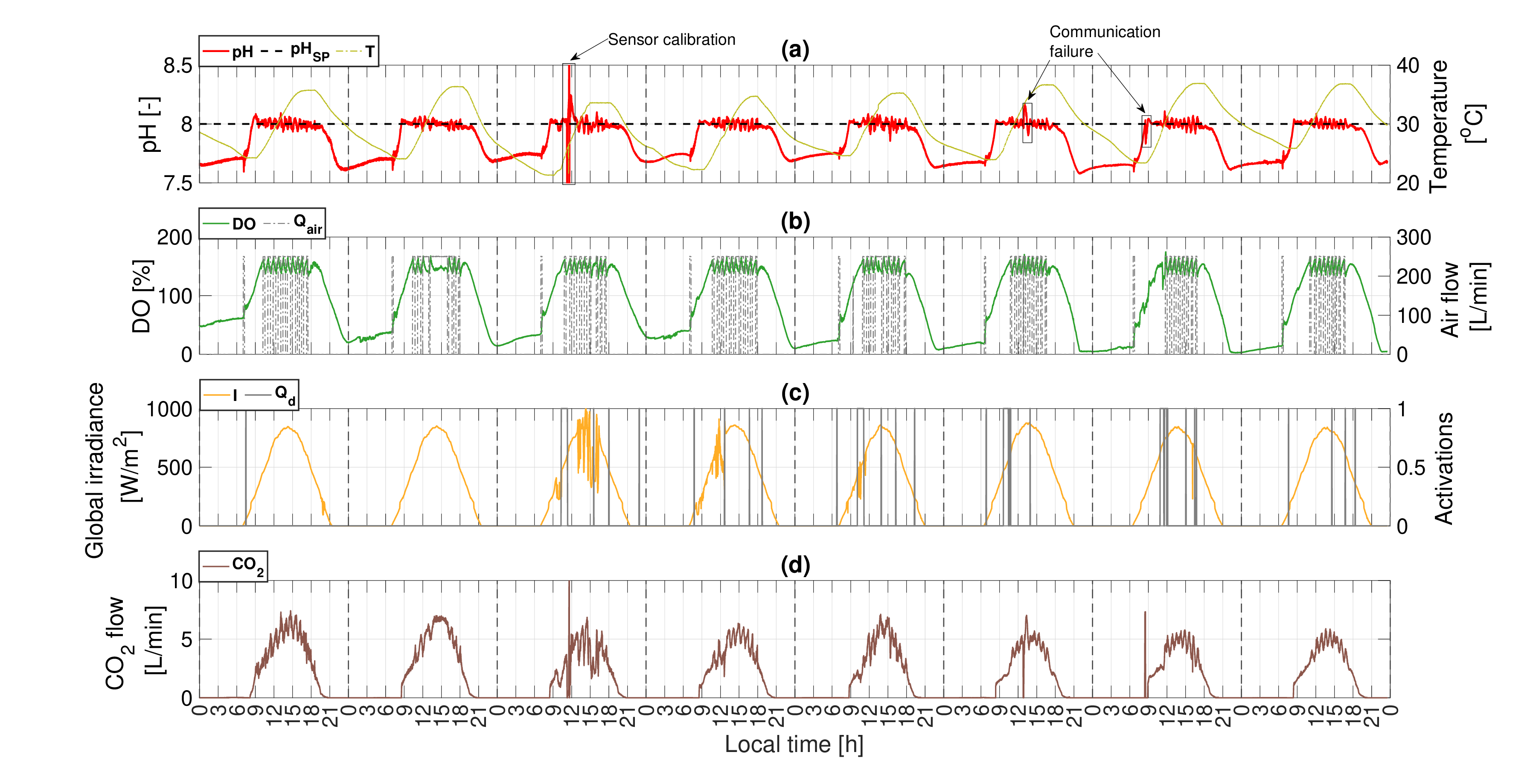}\\
  \caption{Operation of the real PBR system using the proposed methodology. (a) pH in the PBR system (pH), pH reference (pH$_{\mathrm{SP}}$), and PBR system temperature (T); (b) Dissolved oxygen (DO) and air flow injections (Q$_{\mathrm{air}}$); (c) Global irradiance (I), and dilution flow injections (Q$_\mathrm{d}$); (d) CO$_2$ flow (CO$_2$).}\label{fig14}
\end{figure}
\end{landscape}

Once the agent was trained offline, it was deployed on the real system for eight consecutive days, from June 21 to 28, 2025. The results are shown in Fig.~\ref{fig14}. Overall, it can be seen that the agent's control performance was accurate throughout all days (see Fig.~\ref{fig14}~(a)). Note that in these days the temperature and maximum radiation levels were higher than those during the training days. This changed the operating requirements, requiring for example more periodic air injections to maintain the DO at the desired level, which significantly affect the pH control loop. Moreover, in addition to compensating for all these measurable disturbances, the agent also handled several routine operational issues. For instance, on day 3 of operation, the pH sensors were recalibrated by the plant technicians, and on days 6 and 7, there were communication losses between the OPC server and the real PBR facility, which prevented the agent's control actions from being executed in the physical system.

Additionally, the impact of fine-tuning is evident when comparing the different days of operation. For instance, the first two days corresponded to a weekend, during which, as previously mentioned, no harvesting was performed. Although the operating conditions on both days were quite similar, a clear improvement in the agent’s performance was observed on the second day following fine-tuning. Notably, an overshoot in pH relative to the reference occurred at the beginning of the first day, but this was significantly reduced on the second day (see Fig.\ref{fig14}-(a)). Moreover, the control signal on the second day exhibited fewer fluctuations (see Fig.\ref{fig14}-(d)), and the reference tracking was more accurate overall.

Regarding disturbance rejection, during operation days three to five, there were some fluctuations in solar radiation (see Fig.\ref{fig14}-(c)), which were quickly rejected by the controller, resulting in minimal deviations from the reference. The maximum deviation observed due to radiation disturbances—excluding the sensor recalibration and communication failure events—was 0.09 [–], and it occurred on the third day of operation at around 15:00~h (see Fig.\ref{fig14}-(a)). Then, from the fifth day onward, constant dilution flow injections were introduced (see Fig.\ref{fig14}-(c)) due to harvesting activities, as well as increased evaporation caused by higher temperatures (see Fig.\ref{fig14}-(a)). These factors, along with variations in air flow injection (see Fig.\ref{fig14}-(b)), increased the complexity of the control problem. Moreover, these were conditions that did not occur during the training days, which further challenged the performance of the controller. Consequently, especially during the initial days, greater deviations from the reference were observed as a result of these effects. The largest deviation was 0.11 [–] (see Fig.\ref{fig14}-(a)), and occurred on the fifth day at around 15:10 h, caused by a prolonged dilution flow injection (see Fig.\ref{fig14}-(c)). Nevertheless, the fine-tuning process allowed the agent to improve its policy in response to such disturbances, which was particularly evident during the last three days of operation, when the maximum deviation occurred on day 7 at 12:00~h with a value of 0.08 [–].

\section{Conclusions and future works}
\label{section5}
This work proposes an offline–online RL-based control system for pH regulation in microalgae open PBR systems. The RL agent was able to learn a policy based on expert knowledge while also capturing the effects of disturbances through a carefully designed observation space. Through daily online fine-tuning, the agent adapted its control strategy to previously unseen operating conditions, progressively improving its performance over time. The obtained results allow us to draw the following conclusions:

\begin{itemize}
    \item The RL agent learned the basic dynamic behavior of the system during offline training, which was facilitated by the proposed observation space and the applied BC strategy. The utilization of data from a PID-controlled system proved particularly beneficial, as it enabled the initial policy to inherit the performance qualities of these controllers. This foundational understanding was then significantly complemented by the agent's capacity to learn the effect of measurable disturbances, thanks to its comprehensive observation field. 
    
    \item Fine-tuning enabled the transition from a fixed offline-trained policy to an adaptive one suited to new operating conditions. This results fundamental in open PBR systems exposed to changing environmental conditions and the inherent non-linear dynamics of bioprocesses, making adaptive control strategies essential for  efficient operation.

    \item The simulation study highlighted significant benefits. The proposed methodology led to an 8\% improvement in IAE over the PID strategy and a 5\% improvement compared to standard \textit{off-policy} RL controller. Even more notably, control effort showed a substantial reduction, being 54\% lower than PID and 7\% lower than standard  \textit{off-policy} RL controller. This efficiency stems from the system's anticipatory control actions and adaptability, which is particularly important considering the direct impact of control movements on operating costs.
    
    \item Finally, the algorithm's efficacy was validated through an 8-day operational test period in the real system. This real-world evaluation encompassed a range of operational conditions, including both measurable and unmeasurable disturbances inherent to the system's normal functioning, thereby validating the proposed strategy's reliability.
\end{itemize}

Overall, the methodology proposed and followed in this work paves the way for a wider adoption of machine learning algorithms in bioprocess control. Additionally, these methods can also be applied to similar systems with nonlinear dynamics and multiple disturbances.

Future work will focus on enabling the algorithm to automatically incorporate changes in the process reference. This will allow its integration into hierarchical control structures, where these references originate from higher-level optimizations and can therefore be dynamic. Furthermore, future work will focus also on extending the methodology to enable multivariable control of both pH and DO simultaneously.

\section*{Acknowledgements}
This publication is part of the R\&D\&I project PID2023-150739OB-I00, funded by MCIN/ AEI/10.13039/501100011033/ and ``FEDER A way to make Europe", and also by the European Union (Grant agreement ID: 101060991, REALM). In addition, it has funding from the Spanish Ministry of Science, Innovation and Universities through the mobility stay program in foreign higher education and research centers, and from the project for the Strengthening of Research Groups (code P\_FORT\_GRUPOS\_2023/24 - University of Almeria), funded by the General Secretariat for Research and Innovation of the Department of University, Research and Innovation of the Andalusian Regional Government, within the framework of the Andalusia ERDF Operational Programme 2021-2027.

\bibliographystyle{elsarticle-harv} 
\bibliography{biblio}

\end{document}